# An Adaptive Dynamic Replacement Approach for a Multicast based Popularity Aware Prefix Cache Memory System

[1]P. Jayarekha, [2]T.R.Gopalakrishnan Nair
[1]Research Scholar, Dr. MGR University, Dept. of ISE,
BMSCE, Bangalore, jayarekha2001@yahoo.co.in
[2]Director –Research and Industry, Senior Member IEEE, trgnair@ieee.org
Dayananda Sagar Institutions, Bangalore.

Abstract
In this paper we have proposed an adaptive dynamic cache replacement algorithm for a multimedia servers cache system. The goal is to achieve an effective utilization of the cache memory which stores the prefix of popular videos. A replacement policy is usually evaluated using hit ratio, the frequency with which any video is requested. Usually discarding the least recently used page is the policy of choice in cache management. The adaptive dynamic replacement approach for prefix cache is a self tuning, low overhead algorithm that responds online to changing access patterns. It constantly balances between lru and lfu to improve combined result. It automatically adapts to evolving workloads. Since in our algorithm we have considered a prefix caching with multicast transmission of popular objects it utilizes the hard disk and network bandwidth efficiently and increases the number of requests being served.

## 1. INTRODUCTION

A multimedia communication system requires the integration of storage, communication and presentation mechanisms for diverse data types including text, images, audio and video. The reason why multimedia data processing is difficult is, the multimedia servers should possess the following capabilities.

### 1.1 Streaming multiple data streams

A multimedia object can consist of text, audio, video and image data. They differ in storage space and retrieval rate requirements. A multimedia server adds a new dimension to the mechanisms used to store, retrieve, and manipulate the data.

### 1.2 Real-time requirements

Video and audio data are characterized by the fact that they must be presented to the user, and hence retrieved and transported, in real-time.

### 1.3 Large data size

Compared with a typical text object, the size of video or audio object is very large. For example, a 2 hour long movie stored in MPEG-1 format requires over 1 Gbyte of storage.

To provide online services over high speed networks, a multimedia server should be capable of managing efficiently and simultaneously store, access retrieval and transmission of a variety of heterogeneous multimedia objects. In high-performance computer systems, memory bandwidth is often a bottleneck. Minimizing the bandwidth and reducing access latencies are the two primary considerations in the design of multimedia streaming server architecture. A method of prefix caching with multicast transmission of popular objects uses the hard disk and network bandwidth efficiently and increases the number of requests being served. Some of the clips can be easily accommodated in the cache since the modern multimedia servers come with huge cache. Also, it might be very beneficial to store the prefix of the video streams that are getting accessed frequently in the cache. Recent studies have found that nearly 90 percent of media playbacks are terminated prematurely by clients after watching the initial portion of the video. As a result while replacing the videos on the cache, the blocks storing later portions are considered more eligible for replacement than the one storing only the prefix, increasing the hit ratio for the first time requested videos. The choice of the cache organization can have a significant impact on the cost and cache performance. A cache has a fixed amount of storage. In this paper, we have assumed more priority and space to the blocks holding the prefix than the suffix. Whenever this storage space fills, the cache must choose a set of victim blocks to evict and make room for newly requested objects/blocks. The cache replacement policy is one of the factors that determine the effectiveness of cache memories. In general, a replacement policy specifies which block should be removed when a new block must be entered into an already full cache, and should be chosen so as to ensure that blocks likely to be referenced in the near future are retained in the cache.

The replacement policy's goal is to make the best use of the available resources, including disk, memory space and network bandwidth. Our research attempts to improve the performance of the multimedia servers by developing a dynamic replacement strategy that looks at the information available (reference history, access frequency) to make the decision, which block has to be replaced without a proportional increase in the space/time requirements.

A Video-on-Demand (VoD) service offer a large selection of videos which customers can choose. The objective of the VoD system designers is to achieve low access latency for customers. As we know, the disadvantage of unicast VoD is the huge consumption of the server and network bandwidth, so people pay increasingly more attention to multicast VoD [4]. To achieve this allow the server to batch clients requesting the same video and to serve clients in the same batch with one multicast video stream. Another is to give more importance to prefix and evict the suffix of a video while giving room for only prefix. This approach has the advantage to save server resources as well as server access and network bandwidth, thus allowing the server to handle a large number of customers without sacrificing access latency. In this paper, we have proposed an adaptive dynamic replacement algorithm, which balances between LRU and LFU. It combines these two classical methods resulting in a good hit ratio and reducing the waiting time.

The organization of the rest of the paper is as follows: In section 2 we present different replacement policy problem formulation. In Section 3 we discuss about related work. Section 4 presents an adaptive dynamic replacement policy. Section 5 presents algorithm and simulation results. Finally in section 6, we conclude the paper and refer to further work.

## 2. REPLACEMENT POLICY PROBLEM

A cache performs computations on data faster than the data that can be retrieved from the main memory. It attempts to accommodate the data rate at the CPU's demand rate. Three basic cache organizations have been defined at the level of cache memory [1, 2].

### 2.1 Direct Mapped

Each memory block is mapped to a unique cache block. This results in cache block repeatedly evicted even when there are empty slots. In this case, we do not require a replacement mechanism.

### 2.2 Fully Associative

A memory block is mapped to any one of the empty cache blocks, if one exists. If there is no empty cache block, a replacement policy is used to select one of the cache blocks for replacement.

### 2.3 Set Associative

This is a combination of fully-associative and direct-mapped schemes. Cache blocks are grouped into sets, finding a set is like the direct-mapped policy. Finding a block for replacement within the set is like the fully-associative policy. Hence, it is a compromise between the direct-mapped and fully associative placement policies. In general, this organization offers a good balance between hit ratios and implementation costs.

The replacement policy determines how a memory block is mapped to a cache block. While choosing the blocks for replacement, it should ensure that the blocks likely to be referenced in the near future are retained in the cache. The replacement policy has considerable impact on the overall system performance in the selection process of a victim block in fully associative and set associative caches.

Temporal locality refers to two accesses to a block of cache within a small period of time. The shorter the time between the first and last access to cache block the less likely it should be loaded from main memory. The optimization is brought by re-use of the block which has been brought to cache as often as possible.

Cache replacement policies are used to optimize cache management. These policies are used to decide which item to keep and which to discard to make room for the new block. Some of the common replacement policies used are

- Least Recently Used (LRU)
- Least Frequently used(LFU)
- Belady's Min

**Least Recently Used (LRU**): Replaces the block in the cache that has not been used for the longest period of time. From the basics of temporal locality, the blocks that have been referenced in recent past will likely be referenced in the near future. This policy works well when there is a high temporal locality of references in the workload. A Early Eviction LRU (EELRU) proposed in [3], evicts the blocks when it notes that too many pages are being touched in a roughly cyclic pattern that is larger than the main memory.

**Least Frequently Used (LFU)**: It is based on the frequency with which a block is accessed. LFU requires a references count be maintained for each block in the cache. A block- referenced count is incremented by one with each reference . When a replacement is necessary, the LFU replaces/evicts the blocks/objects with the lowest reference count.

**Belady's Min:** The most efficient caching algorithm would be always discarding the information that will not be needed for the longest time in the future. This optimal result is referred to as Belady's optimal algorithm or the clairvoyant algorithm. This policy is generally not implementable since it is impossible to predict how far in the future the information will be needed.

## 3. RELATED WORK

**Least Frequently Used (LFU)-Aging**: The LFU [5] policy can suffer from cache pollution (an effect of temporal locality): if a formerly popular object becomes unpopular, it will remain in the cache for a long time, preventing other newly or slightly less popular objects from replacing it. The aging policy is applied at intervals to bring down the reference counts of such objects and ultimately make them candidates for replacement. The LFU-Aging policy is similar to the LFU policy. The least frequently used document is replaced when space is required for a new document. However, LFU-Aging attempts to deal with the problem of LFU. With LFU, some blocks can build up extremely high reference counts so that they are rarely (if ever) replaced, even if these blocks are never requested again. The LFU-Aging policy attempts to ensure that this condition does not occur by limiting and aging (i.e., occasionally reducing) reference counts. LFU-Aging addresses cache pollution when it considers both block access frequency and its age in cache. There is a variant called LFU with Dynamic Aging (LFUDA) that uses dynamic aging to accommodate shifts in the set of popular objects. It adds a cache age factor to the reference count when a new object is added to the cache or when an existing object is re-referenced. LFUDA increments the cache ages when evicting blocks by setting it to the evicted object's key value. Thus, the cache age is always less than or equal to the minimum key value in the cache. In our approach a similar dynamic aging factor, not only helps in evicting a video which is currently offline, but also in balancing the frequency and time interval between the requests.

**Greedy Dual Size (GDS)**: It combines temporal locality, size, and other cost information. The algorithm assigns a cost/size value to each cache block [13]. In the simplest case the cost is set to 1 to maximize the hit ratio, but costs such as latency, network bandwidth can be explored. GDS assigns a key value to each object. The key is computed as the objects reference count plus the cost information divided by its size. The algorithm takes into account recency for a block by inflating the key value (cost/size value) for an accessed block by the least value of currently cached blocks. The GDS-aging version adds the cache age factor to the key factor. By adding the cache age factor, it limits the influence of previously popular documents.

**Frequency Based Replacement (FBR)**: This is a hybrid replacement policy, attempting to capture the benefits of both LRU and LFU without the associated drawbacks [14]. FBR maintains the LRU ordering of all blocks in the cache, but the replacement decision is primarily based upon the frequency count. To accomplish this, FBR divides the cache into three partitions: a new partition, a middle partition and an old partition. The new partition contains the most recent used blocks (MRU) and the old partition the LRU blocks. The middle section consists of those blocks neither in the new or old section. When a reference occurs to a block in the new section, its reference count is not incremented. References to the middle and old sections cause the reference counts to be incremented. When a block must be chosen for replacement, FBR chooses the block with the lowest reference count, but only among those blocks that are in the old section.

**Random (RAND)**: It chooses among all blocks in the cache with equal probability. Intuitively, RAND seems appealing in this context if the client's caches are filtering all the locality characteristics from their reference streams. RAND provides a kind of lower bound on performance that is, there is no reason to use any policy that performs worse than RAND. Thus, this policy must be used if it is faster and less expensive.

**Prediction**: An optimal cache replacement policy would know a document's future popularity and choose the most advantageous way to use its finite space. Unfortunately this requires future knowledge, and even with perfect future knowledge it is still computationally expensive.

In general, the policies anticipate future memory references by looking at the past behavior of the programs (program's memory access patterns). Their job is to identify a block (containing memory references) which should be thrown away, in order to make room for the newly referred line that experienced a miss in the cache. Relative performance of these algorithms depends mainly on the length of the history consulted, but they ignore the cache block state information that is also indicative of the characteristics of the program.

For the purpose of maximizing the hit ratio, increasing the number of videos getting serviced and to maximize the average time between successive page replacements, an efficient replacement algorithm is proposed which uses Access Interval Predictor (AIP) [10] and Greedy Dual size approach.

In our approach, we have considered to store the prefix, which is assumed to be of equal size. We have added an Age factor to list $L_2$, which avoids the cache pollution. But authors [8, 11, 12] have not discussed about storing the popular prefix and transmitting it through a multicast group. While replacing any video we have not only considered the frequency, but also, since how long that video is there in the cache. A video which might have been popular some time ago, but which has not been requested for some time interval will be made to evict from the cache. Hence, the decision on which prefix to be evicted is taken dynamically depending on the time and the current popularity of that video. The deadline of any multicast group is dependent on the deadline time of the first client joining the group.

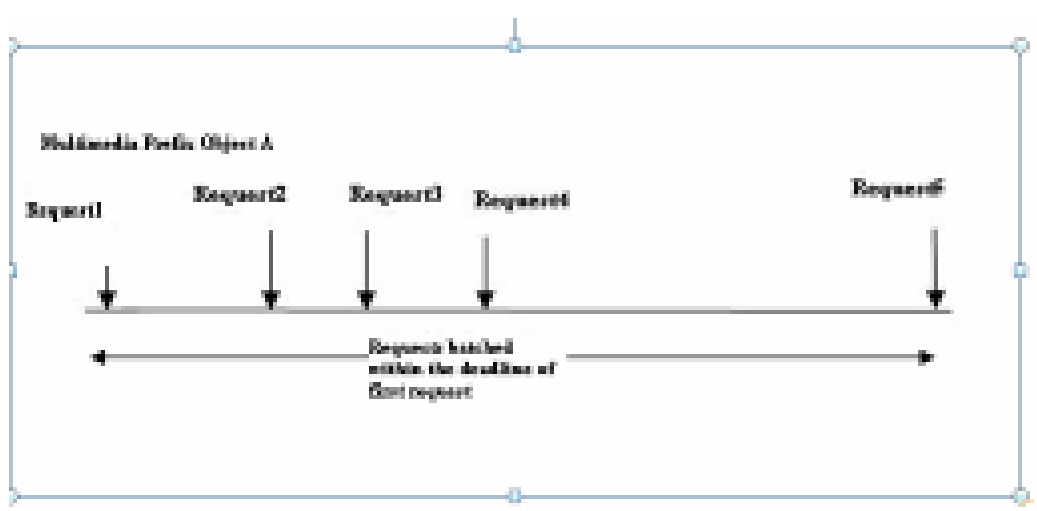

Figure 1. Batching of request to a multicast group

The popularity is defined as the number of clients joining to that group. This is saved as one of the parameters with the prefix of the video. This helps in studying the history of any video.

## 4. AN ADAPTIVE DYNAMIC REPLACEMENT APPROACH

This policy maintains two LRU page lists for the prefix that is being stored in the cache. $L_1$ list maintains the prefix that has been requested only once recently, while $L_2$ maintains that has been requested at least twice recently [8]. The prefix of the video which is being watched many times in a short interval of time is thought of having high frequency. Hence $L_1$ stores the level of being recent, while $L_2$ maintains the frequency. Each prefix in the list $L_1$ or $L_2$ belong to a multicast group. The frequency of any prefix denotes the total number of clients currently being streamed from that group concurrently within the deadline limits. A deadline for any group depends on the deadline limit of the first member joining that group.

These two lists are extended with a ghost list ($B_1$ or $B_2$) which are attached to the bottom of the two lists. Ghost list is used to keep track of the history of recently evicted cache entries. The ghost list contains only the Meta data not the data itself, as an entry is evicted into a ghost list its data is discarded. $B_1$ holds the entries evicted from the list $L_1$ but are still tracked. Similarly $B_2$ has the entries evicted from the list $L_2$.

Let $c$ be the cache size in pages. We have a imaginary cache of size 2c. A prefix of a video is said to be offline if it has finished streaming all the requests of that multicast group. It is said to be online if, it can accept a new member to a multicast group within the limits of the deadline of that group. Adaptive Replacement policy maintains two LRU lists. $L_1$ that contain pages that have been seen recently only once and $L_2$ that contain pages that have been seen at least twice recently. More precisely, a page resides in $L_1$ if it has been requested exactly once and not evicted and moved to $B_1$. Similarly, a page resides in $L_2$ if it has been requested more than once since the last time it was removed was from $L_2$ moved to $B_2$ .

The policy functions are as follows: If $L_1$ contains exactly $c$ pages, replace the LRU page in $L_1$, move the meta data of that page to B1.Otherwise, replace the LRU page in $L_2$. Initially, the lists are empty: $L_1 = L_2 = \Phi$. If a requested page resides in $L_1$ or $L_2$, the policy moves it to the MRU position of $L_2$. Otherwise, it moves to the MRU position of $L_1$. In the latter case, if $|L_1| + |B_1| < c$, then the policy removes the LRU member of $L_1$ and if the requested page resides in the cache, but it is offline, it has finished streaming all the requests within the deadline limits of that multicast group. The requested page is made online, by moving it to the MRU position of $L_2$.

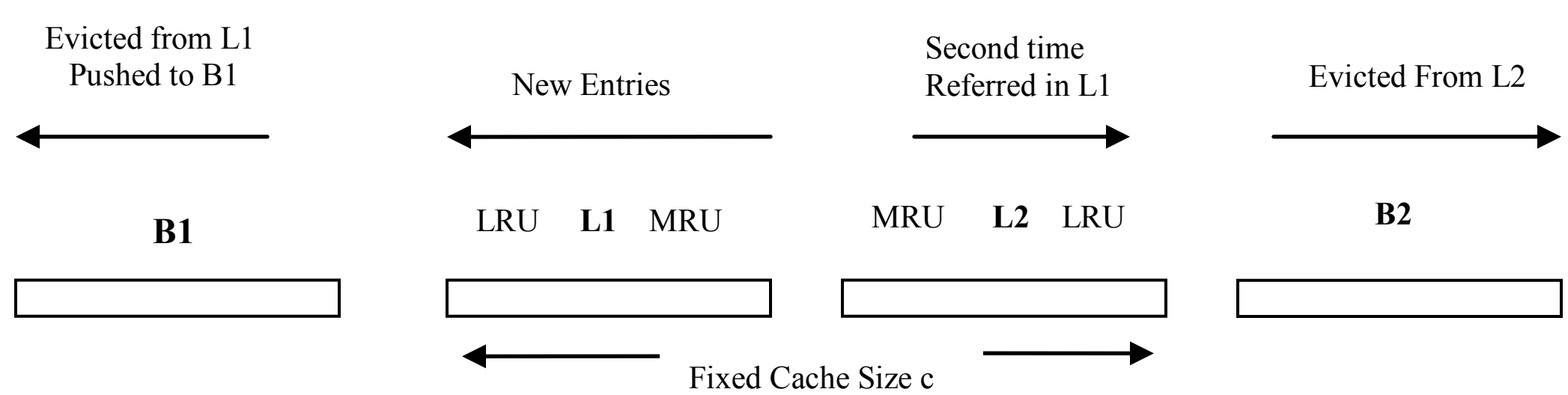

Figure 2. An Adaptive Dynamic replacement System

Hits in $B_1$ will increase the size of $L_1$, if it exceeds $c$, the last entry of $L_2$.is evicted and moved to $B_2$.

Hits in $B_2$ will shrink $L_1$, the last entry of $L_1$ is now evicted into $B_1$.

## 5. ALGORITHM AND SIMULATION RESULTS

**Algorithm**
**Nomenclature**

L1: List maintains the prefix that has been requested recently only once.
$L_2$: List maintains the prefix that has been requested at least twice recently.
$B_1$: Ghost list of $L_1$, recently evicted from list $L_1$
$B_2$: Ghost list of $L_2$, recently evicted from list $L_2$.
c : Number of pages in the cache.

Initialize $L_1, B_1, L_2, B_2 = \Phi$
*Let* $x$ be the requested page
If cache is not full and $x$ is on-line
**Case 1**: $x \varepsilon L_2$ Increase the frequency count of $x$ and Timestamp of all the pages of $L_2$.
**Case 2:** $x \varepsilon B_1$ move $x$ to the MRU of $L_2$.
**Case 3:** $x \varepsilon B_2$ move $x$ to the MRU of $L_1$.
**Case 4**: If $x$ is not present,
If $|B_1| + |L_1| <= c$
Replace the LRU page in $L_1$ by evicting LRU page and moving it to $B_1$.
Else
Replace the page with least age factor of $L_2$ moving it to $B_2$.
**Case 5:** End

The factors considered while selecting a victim for replacement in $L_2$ is as follows:

a. All the pages are of equal size, since only the prefixes of different videos are stored.
b. The frequency for each prefix represents the number of requests serviced in a multicast group.
c. Each time a request occurs, the time at which the request arrived is compared with the time stamp of the prefix. If time stamp is greater than deadline time of a multicast group, the prefix is said to be offline, hence a new request cannot be added to it. The request can be only serviced as a new request.
d. A request is said to be offline if it has completed servicing all the requests within the deadline limits.
e. An inflation factor is added to the offline prefix. This factor makes the Age-factor of the prefix closer to itself to become a candidate for replacement.

We have considered the sample example of simulation. All the prefixes are assumed to be online.

The first row represents the pages present currently in $L_2$.The pages are arranged in the increasing order of timestamp, how recently the pages are requested. Frequency count tells the popularity of that video. When a page has to be evicted from $L_2$ and to give the room for the new one, the following score-value is used.

$$\textbf{Scorevalue} = \frac{(\text{Frequecy})}{(\text{Time Stamp})}$$

Table 1.

| MRU | | $L_2$ | | LRU | |
|---|---|---|---|---|---|
| Page | E | A | D | F | S |
| Time stamp | 1 | 03 | 10 | 15 | 17 |
| Frequency | 2 | 4 | 2 | 6 | 13 |

As the timestamp increases, the Score-factor decreases. The one with the least Score-factor will be the candidate for replacement. If the page was most popular, it will remain in the cache resulting in cache pollution. Always, the least frequently used page will not be made as a victim for replacement. Instead a most popular page will also become a candidate for replacement if the page is not requested for a very long time.

As shown in table.1, page D will be the candidate replacement since it has the least score_value. A new page G is added at the MRU position as shown in table 2.

Table 2

| MRU | | $L_2$ | | LRU | |
|---|---|---|---|---|---|
| Page | G | F | A | F | S |
| Time stamp | 1 | 04 | 11 | 16 | 18 |
| Frequency | 3 | 4 | 2 | 6 | 13 |

If a request for page A which is present in the cache occurs, then the changes observed in $L_2$ list is as shown in table 3.

Table 3

| MRU | | L2 | | LRU | |
|---|---|---|---|---|---|
| Page | A | G | F | F | S |
| Time stamp | 1 | 02 | 05 | 17 | 19 |
| Frequency | 3 | 3 | 4 | 6 | 13 |

Suppose at time stamp 18, prefix page S becomes offline, the frequency is reduced by 10%. This works as an inflation factor reducing overall value of score-value. If this prefix has not been requested once again over a long interval of time, the inflation factor is applied again so that this page is eligible for replacement.

**Simulation Results**

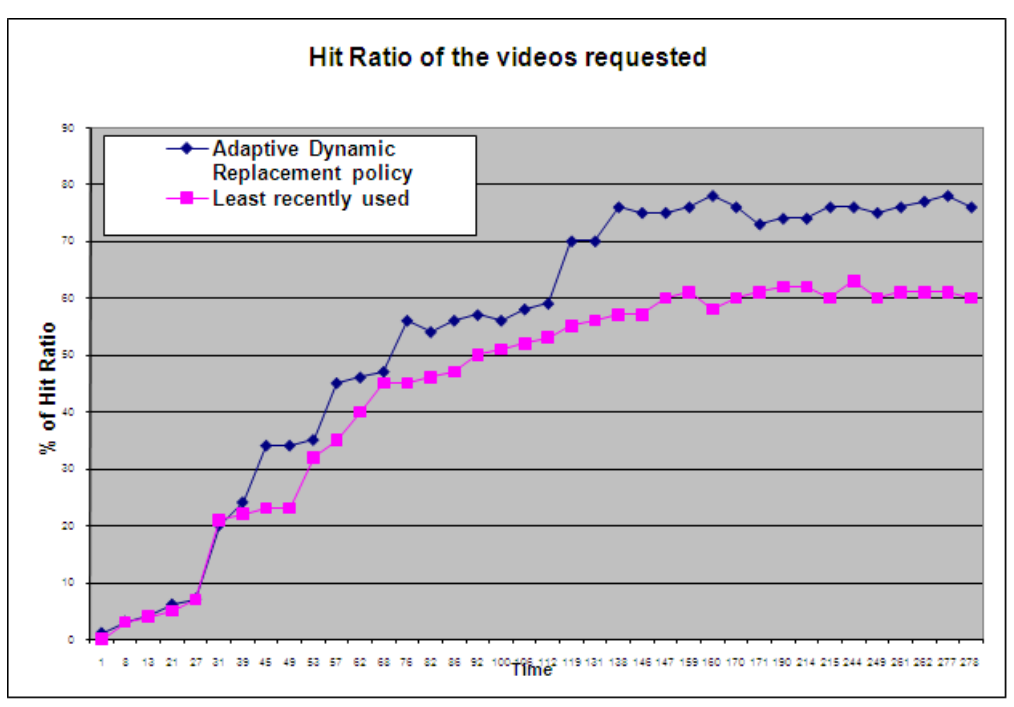

Figure 3. Hit Ratio of the videos requested

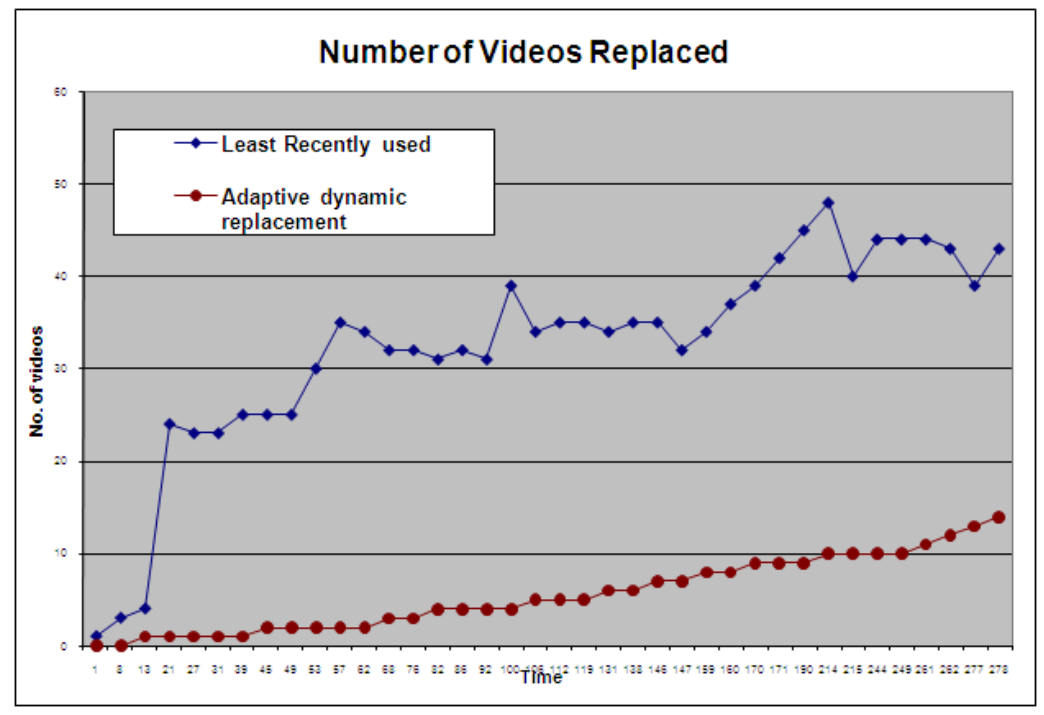

Figure 4. Number of Videos Replaced

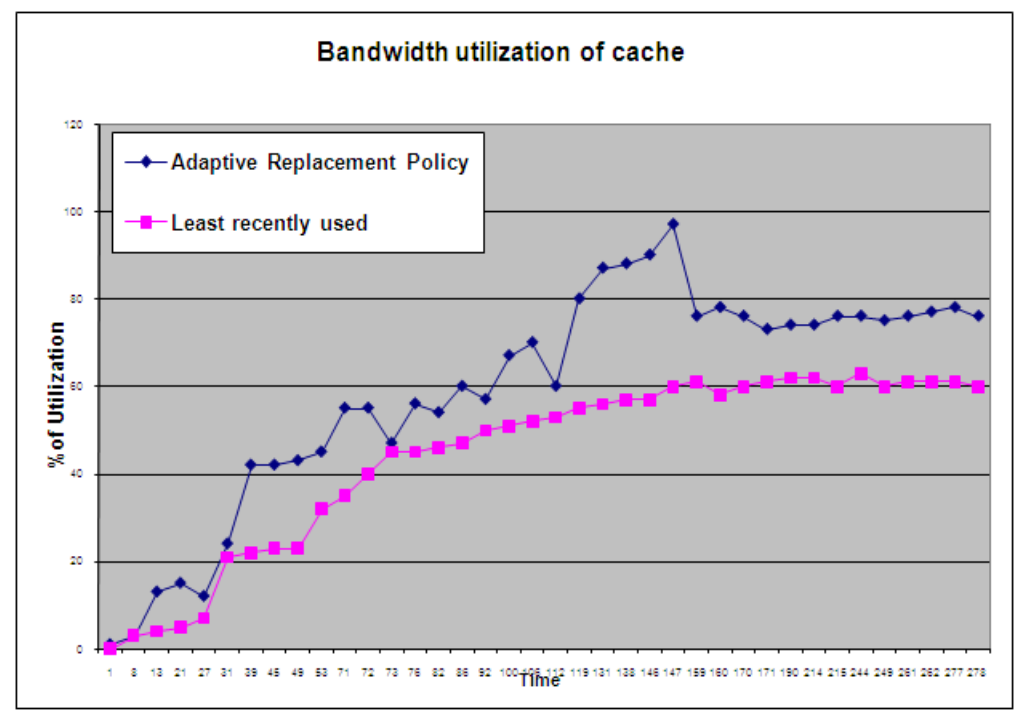

Figure 5. Bandwidth utilization of cache

Our simulation model consists of two lists one to store $L_1$ which stores the recency, while $L_2$ maintains the frequency. $B_1$ and $B_2$ holds the pages evicted from the $L_1$and $L_2$ respectively. Fig 2 shows the results in terms of hit ratio. Our algorithm has more hit ratio compared with LRU. Since the evicted pages from the list L1 and L2 is as shown in Fig 3, we have the evicted pages stored in $B_1$ and $B_2$.

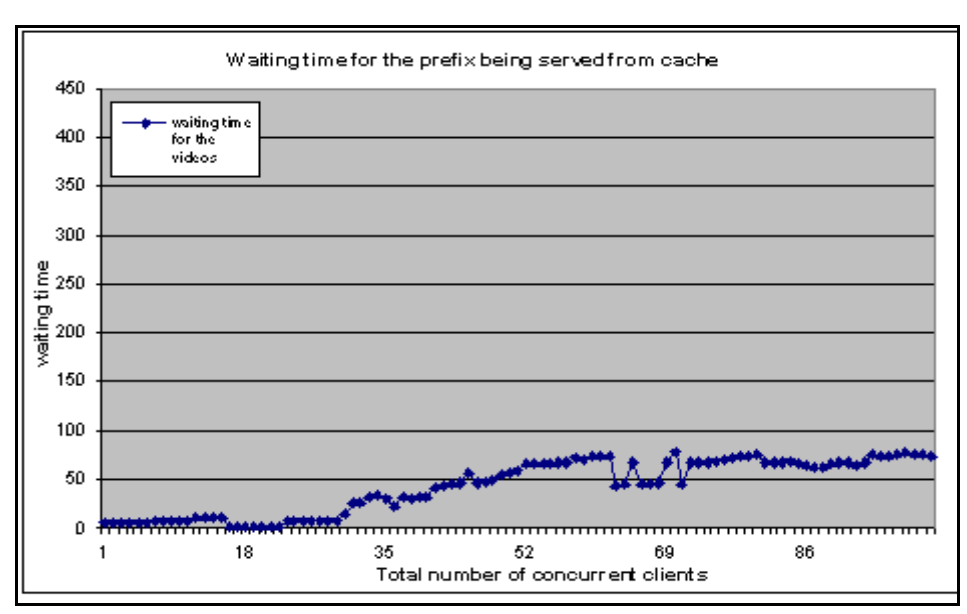

Figure 6. Waiting time for the prefix being served from cache

The number of replacements is also reduced. In Fig 4 our algorithm has shown that, compared with LRU the percentage of utilization of cache bandwidth is more. Fig 5 proves the reduction in waiting time.

## 6. CONCLUSION

In this paper we have proposed an adaptive dynamic replacement algorithm based on prefix caching. Our multimedia architecture supports multicast transmission so that it increases the number of clients being serviced.

From the simulation results it is shown that our algorithm works better than LRU by maintaining the list of recently evicted videos.

However, in this algorithm by default always LRU page is evicted, but we can further improve the results by predictive prefetching any page depending on some weight factor, which decides whether the page can be moved into the active list.

## REFERENCES

[1] Cho S, King J, Lee G, "Coherence and Replacement Protocol of DICE-A Bus Based COMA Multiprocessor", Journal of Parallel and Distributed Computing, Vol. 57 pp. 14-32, 1999.

[2] Aguilar J, Leiss E, "A Proposal for a Consistent Framework of Dynamic/Adaptive Policies for Cache Memory Management" Technical Report, Department of Computer Science, University of Houston, April2000.

[3] Smaradakis Y, Kaplan S, Wilson P, "EELRU: Simple and Effective Adaptive Page Replacement Performance Evaluation" Review, Vol. 27, pp. 122-133, January 1999.

[4] S Ramesh, I Rhee, and K Cuo, "Multicast with Cache (Mcache): An Adaptive Zero-Delay Video-on-Demand Service", *IEEE Infocom***,** pp. 85-94, Anchorage, Alaska, USA, April 2001.

[5] Jose Aguilar , Ernst Leiss "A Web Proxy Cache Coherency and Replacement Approach" Springer Berlin Heidelberg Volume 2198/2001 Jan 2001.

[6] K Li, T Nanya, H Shen, F Chin, W Zhang "An efficient cache replacement algorithm for multimedia object caching" International Journal of Computer Systems Science, 2007.

[7] A Wierzbicki, N Leibowitz, M Ripeanu, R Wozniak "Cache replacement policies revisited: The case of P2P traffic" IEEE International Symposium on Cluster Computing 2004.

[8] Nimrod Megiddo , Dharmendra S. Modha, "Outperforming LRU with an Adaptive Replacement Cache Algorithm**"** IEEE computer Society Press 2004.

[9] K Li, H Shen, K Tajima, L Huang "An effective cache replacement algorithm in transcoding-enabled proxies" The Journal of Supercomputing, Springer 2006.

[10] M Kharbutli, Y Solihin – "Counter-based cache replacement algorithms" IEEE International Conference on Computer Design VLSI 2005.

[11] Z Li, D Liu and H Bi "CRFP: A Novel Adaptive Replacement Policy Combined the LRU and LFU Policies" 2008 IEEE 8th International Conference on Computer and Information Technology Workshops.

[12] Y smaragdakis, "General adaptive replacement policies" proceedings of the 4th international symposium on Memory 2004.

[13] Shudong Jin **"**Popularity-Aware Greedy Dual-Size Web Proxy Caching Algorithms" The 20th International Conference on Distributed Computing Systems ICDCS 2000.

[14] John T Robinson, "Data cache management using frequency-based replacement", ACM SIGMETRICS, 1990

## BIOGRAPHY

**P Jayarekha** holds M.Tech in Computer Science securing second rank . She has fifteen years experience in teaching field. She has published many papers. Currently she is working as a teaching faculty in the department of Information science and engineering at BMS College of Engineering, Bangalore ,India.

**T.R. Gopalakrishnan Nair** holds M.Tech. (IISc, Bangalore) and Ph.D. degree in Computer Science. He has 3 decades experience in Computer Science and Engineering through research, industry and education. He has published several papers and holds patents in multi domains. He won the PARAM Award for technology innovation. Currently he is the Director of Research and Industry in Dayananda Sagar Institutions, Bangalore, India.